\documentclass[12pt]{article}

\usepackage{graphicx}

\begin{document}

\begin{center}
{\bf Inflation of Universe by Nonlinear Electrodynamics} \\
\vspace{5mm} S. I. Kruglov
\footnote{E-mail: serguei.krouglov@utoronto.ca}

\vspace{3mm}
\textit{Department of Physics, University of Toronto, \\60 St. Georges St.,
Toronto, ON M5S 1A7, Canada\\
Department of Chemical and Physical Sciences, University of Toronto,\\
3359 Mississauga Road North, Mississauga, ON L5L 1C6, Canada} \\
\vspace{5mm}
\end{center}

\begin{abstract}
Nonlinear electrodynamics with two dimensional parameters is studied. The range of electromagnetic fields when principles of causality, unitarity and the classical stability hold, are obtained. A singularity of the electric field at the center of charges is absent within our model and there are corrections to the Coulomb law as $r\rightarrow\infty$. The universe inflation takes place in the background of stochastic magnetic fields. The second stage of the universe evolution is the radiation era so that the graceful exit exists. We estimated the spectral index, the tensor-to-scalar ratio, and the running of the spectral index that are in a rough accord with the PLANCK and WMAP data.
\end{abstract}

\section{Introduction}

The inflation of the universe can be described by coupling GR with nonlinear electromagnetic fields \cite{Garcia}-\cite{Kruglov5}. In the early time of the universe evolution electromagnetic fields were very strong and quantum corrections should be taken into account \cite{Jackson} and, as a result, Maxwell's electrodynamics becomes nonlinear electrodynamics (NED) \cite{Heisenberg}, \cite{Schwinger}, \cite{Adler}.  First NED, without singularities of point-like charges, was proposed by Born and Infeld \cite{Born}.
In this paper we study NED that for weak fields leads to the Maxwell limit. We assume that the universe filled by the stochastic magnetic background. Stochastic fluctuations in electron-positron plasma can lead to a stochastic magnetic field  \cite{Lemoine}, \cite{Lemoine1}. We consider the magnetic background because the electric field is screened by the charged primordial plasma \cite{Lemoine1}.

In Sect. 2 the causality and unitarity principles in a NED model with two dimensional parameters $\beta$, $\gamma$ are investigated. We study field equations and their dual invariance in Sect. 3. It is shown that there is no singularity of the electric field at the origin of the point-like charges and the electric field possesses the maximum. There are corrections to Coulomb's law in the order of ${\cal O}(r^{-6})$. We estimate the model parameters $\beta$ and $\gamma$ by the requirement that at the weak field limit our model is converted into QED with one loop correction. In Sect. 4 we investigate the cosmology of the universe which is filled by stochastic magnetic fields. The energy density and pressure as the functions of the scale factor are obtained. It was demonstrated that there are no the singularity of the Ricci scalar. The evolution of the universe is studied in Sect. 5. The scale factor as a function of the time is obtained. The bound on the speed of sound which guarantees the classical stability and causality is calculated in Sect. 6. In Sect. 7 we evaluate the cosmological parameters.

\section{The NED model}

Here, we propose NED, which is a modification of models \cite{Kruglov4} and \cite{Kruglov5}, with the Lagrangian given by
\begin{equation}
{\cal L} = -\frac{{\cal F}}{(1+2\beta{\cal F})^{3/2}}+\frac{\gamma}{2}{\cal G}^2,
\label{1}
\end{equation}
where ${\cal F}=(1/4)F_{\mu\nu}F^{\mu\nu}=(\textbf{B}^2-\textbf{E}^2)/2$, ${\cal G}=\textbf{B}\cdot\textbf{E}$, and $\beta$ ($\beta>0$), $\gamma$ ($\gamma>0$) are dimensional parameters.
This Lagrangian will lead to the attractive scenario of the universe inflation with the graceful exit and avoiding super-inflationary behavior as in \cite{Kruglov4} and \cite{Kruglov5}. The stress-energy tensor following from Eq. (1) is
\[
T_{\mu\nu}={\cal L}_{\cal F}F_\mu^{~\alpha}F_{\nu\alpha}+\frac{1}{2}{\cal L}_{\cal G}\left(F_\mu^{~\alpha}\tilde{F}_{\nu\alpha}
+F_\nu^{~\alpha}\tilde{F}_{\mu\alpha}\right)-g_{\mu\nu}{\cal L}
\]
\begin{equation}
=\frac{(\beta{\cal F}-1)F_\mu^{~\alpha}F_{\nu\alpha}}{(1+2\beta{\cal F})^{5/2}}+\frac{1}{2}\gamma{\cal G}\left(F_\mu^\alpha\tilde{F}_{\nu\alpha}+F_\nu^{~\alpha}\tilde{F}_{\mu\alpha}\right)-g_{\mu\nu}{\cal L},
\label{2}
\end{equation}
with ${\cal L}_{\cal F}=\partial{\cal L}/\partial{\cal F}$, ${\cal L}_{\cal G}=\partial{\cal L}/\partial{\cal G}$.
Making use of Eq. (2) we find the trace of the stress-energy tensor
\begin{equation}
{\cal T}\equiv T_\mu^{~\mu}=\frac{12\beta{\cal F}^2}{(1+2\beta{\cal F})^{5/2}}+2\gamma {\cal G}^2.
\label{3}
\end{equation}
In linear electrodynamics $\beta\rightarrow 0$, $\gamma\rightarrow 0$ and ${\cal L}\rightarrow -{\cal F}$ so that the stress-energy tensor is traceless, ${\cal T}\rightarrow 0$. Because the dimensional parameters are present in the model the scale invariance is violated and the stress-energy tensor is not traceless.

In accordance with the causality principle the group velocity of excitations over the background should be less than the speed of light and there will be not tachyons in the theory spectrum. The unitarity principle guarantees the absence of ghosts. Both principles lead to the inequalities \cite{Shabad2}:
\[
 {\cal L}_{\cal F}\leq 0,~~~~{\cal L}_{{\cal F}{\cal F}}\geq 0,~~~~{\cal L}_{{\cal G}{\cal G}}\geq 0,
\] \begin{equation}
{\cal L}_{\cal F}+2{\cal F} {\cal L}_{{\cal F}{\cal F}}\leq 0,~~~~2{\cal F} {\cal L}_{{\cal G}{\cal G}}-{\cal L}_{\cal F}\geq 0.
\label{4}
\end{equation}
By virtue of Eq. (1) one obtains
\[
{\cal L}_{\cal F}= \frac{\beta{\cal F}-1}{(1+2\beta{\cal F})^{5/2}},~~~~ {\cal L}_{{\cal G}{\cal G}}=\gamma,~~~~
2{\cal F}{\cal L}_{{\cal G}{\cal G}}-{\cal L}_{{\cal F}}=2{\cal F}\gamma+\frac{1-\beta{\cal F}}{(1+2\beta{\cal F})^{5/2}},
\]
\begin{equation}
{\cal L}_{\cal F}+2{\cal F}{\cal L}_{{\cal F}{\cal F}}=\frac{-4(\beta{\cal F})^2+11\beta{\cal F}-1}{(1+2\beta{\cal F})^{7/2}},~~~~
{\cal L}_{{\cal F}{\cal F}}=\frac{3\beta(2-\beta{\cal F})}{(1+2\beta{\cal F})^{7/2}}.
\label{5}
\end{equation}
Making use of Eqs. (4) and (5), in the case of $\gamma=0$, $\textbf{B}=0$, we obtain $|\textbf{E}|\leq \sqrt{1/\beta}$ which is satisfied because the maximum of the electric field is given by $|\textbf{E}_{max}|= \sqrt{1/\beta}$ (see Eq. (17)).
If $\gamma=0$, $\textbf{E}=0$, one has $|\textbf{B}|\leq \sqrt{(11-\sqrt{105})/(4\beta)}\approx 0.434/\sqrt{\beta}$.

\section{Electromagnetic field equations}

With the help of Eq. (1) we find field equations
\begin{equation}
\partial_\mu\left({\cal L}_{\cal F}F^{\mu\nu} +{\cal L}_{\cal G}\tilde{F}^{\mu\nu} \right)=0.
\label{6}
\end{equation}
Making use of Eqs. (1) and (6) we obtain
\begin{equation}
 \partial_\mu\left(\frac{(\beta{\cal F}-1)F^{\mu\nu}}{(1+2\beta{\cal F})^{5/2}}
+\gamma{\cal G}\tilde{F}^{\mu\nu}\right)=0.
\label{7}
\end{equation}
Using the equation $\textbf{D}=\partial{\cal L}/\partial \textbf{E}$, we find the electric displacement field
\begin{equation}
\textbf{D}=\frac{1-\beta{\cal F}}{(1+2\beta{\cal F})^{5/2}} \textbf{E}+\gamma {\cal G}\textbf{B}.
\label{8}
\end{equation}
The magnetic field $\textbf{H}=-\partial{\cal L}/\partial \textbf{B}$ is given by
\begin{equation}
\textbf{H}= \frac{1-\beta{\cal F}}{(1+2\beta{\cal F})^{5/2}}\textbf{B}-\gamma{\cal G}\textbf{E}.
\label{9}
\end{equation}
We use the decomposition of Eqs. (8) and (9) as follows \cite{Hehl}:
\begin{equation}
D_i=\varepsilon_{ij}E^j+\nu_{ij}B^j,~~~~H_i=(\mu^{-1})_{ij}B^j-\nu_{ji}E^j.
\label{10}
\end{equation}
With the help of Eqs. (8), (9) and (10) we obtain
\[
\varepsilon_{ij}=\delta_{ij}\varepsilon,~~~~(\mu^{-1})_{ij}=\delta_{ij}\mu^{-1},~~~~\nu_{ji}=\delta_{ij}\nu,
\]
\begin{equation}
\varepsilon=\frac{1-\beta{\cal F}}{(1+2\beta{\cal F})^{5/2}},~~~~
\mu^{-1}=\varepsilon=\frac{1-\beta{\cal F}}{(1+2\beta{\cal F})^{5/2}},~~~~\nu=\gamma {\cal G}.
\label{11}
\end{equation}
Field equation (7), by virtue of Eqs. (8) and (9), can be represented as the Maxwell equations
\begin{equation}
\nabla\cdot \textbf{D}= 0,~~~~ \frac{\partial\textbf{D}}{\partial
t}-\nabla\times\textbf{H}=0.
\label{12}
\end{equation}
Using the Bianchi identity $\partial_\mu \tilde{F}^{\mu\nu}=0$, we obtain
\begin{equation}
\nabla\cdot \textbf{B}= 0,~~~~ \frac{\partial\textbf{B}}{\partial
t}+\nabla\times\textbf{E}=0.
\label{13}
\end{equation}
With the aid Eqs. (8) and (9) we find
\begin{equation}
\textbf{D}\cdot\textbf{H}=(\varepsilon^2-\nu^2)\textbf{E}\cdot\textbf{B}+\varepsilon\nu(\textbf{B}^2-\textbf{E}^2).
\label{14}
\end{equation}
The dual symmetry is violated as $\textbf{D}\cdot\textbf{H}\neq\textbf{E}\cdot\textbf{B}$  \cite{Gibbons}. In BI electrodynamics and in Maxwell's electrodynamics ($\varepsilon=1$, $\nu=0$) the dual symmetry occurs. In quantum electrodynamics with quantum corrections and in generalized BI electrodynamics \cite{Krug} the dual symmetry is violated.

When the source is a point-like electric charge $q_e$, the equation for the electric displacement field is
\begin{equation}
\nabla\cdot \textbf{D}=4\pi q_e\delta(\textbf{r}).
\label{15}
\end{equation}
Making use of Eq. (8) at $\textbf{B}=0$ the solution to Eq. (15) is given by
\begin{equation}
\frac{E\left(2+\beta E^2\right)}{2(1-\beta E^2)^{5/2}}=\frac{q_e}{r^2}.
\label{16}
\end{equation}
From Eq. (16) as $r\rightarrow 0$ we obtain the solution
\begin{equation}
E(0)=\sqrt{\frac{1}{\beta}}.
\label{17}
\end{equation}
Equation (17) shows the maximum of the electric field in the center of charged particles similar to BI electrodynamics. Thus, at the origin of the point-like charges there is no singularity of the electric field unlike the Maxwell electrodynamics. Let us introduce unitless variables
\begin{equation}
x=\frac{r^2}{q_e\sqrt{\beta}},~~~~y=\sqrt{\beta}E.
\label{18}
\end{equation}
Then Eq. (16) can be written as follows:
\begin{equation}
\frac{y(2+y^2)}{2(1-y^2)^{5/2}}=\frac{1}{x}.
\label{19}
\end{equation}
The function $y(x)$ is depicted in Fig. 1.
\begin{figure}[h]
\includegraphics[height=3.0in,width=3.0in]{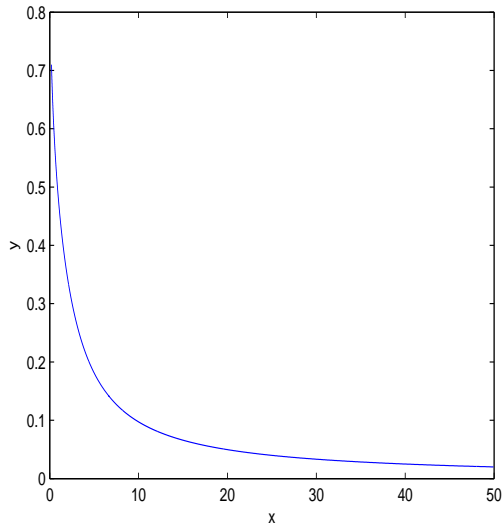}
\caption{\label{fig.1}The function  $y(x)$.}
\end{figure}
The numerical solutions to Eq. (19) are given by Table 1.
\begin{table}[ht]
\caption{}
\centering
\begin{tabular}{c c c c c c c c c c  c}\\[1ex]
\hline
$x$ & 1 & 2 & 3 & 4 & 5 & 6 & 7 & 8 & 9 & 10\\[0.5ex]
\hline
 $y$ & 0.475 & 0.344 & 0.267 & 0.217 & 0.181 & 0.155 & 0.135 & 0.120 & 0.107 & 0.097\\[0.5ex]
\hline
\end{tabular}
\end{table}
The function $y(x)$ as $x\rightarrow \infty$ has the asymptotic
\begin{equation}
y=\frac{1}{x}-\frac{3}{x^3}+{\cal O}(x^{-5}).
\label{20}
\end{equation}
Making use of  Eqs. (18) and (20) we obtain the electric field as $r\rightarrow\infty$
\begin{equation}
E(r)=\frac{q_e}{r^2}-\frac{3\beta q_e^3}{r^6}+{\cal O}(r^{-10}).
\label{21}
\end{equation}
According to Eq. (21) Coulomb's law possesses corrections in the order of ${\cal O}(r^{-6})$. At $\beta=0$  one finds the Coulomb law $E=q_e/r^2$ similar to Maxwell's electrodynamics.
Making use of Eq. (19), one finds the asymptotic of $y(x)$ as $x\rightarrow 0$
\begin{equation}
y(x)=1-0.59x^{0.4}~~~~~~~x\rightarrow 0.
\label{22}
\end{equation}
With the help of Eqs. (18) and (22), we obtain the electric field
\begin{equation}
E(r)=\frac{1}{\sqrt{\beta}}-\frac{0.59r^{0.8}}{q_e^{0.4}\beta^{0.7}}~~~~~~r\rightarrow 0.
\label{23}
\end{equation}
Equation (23) at $r=0$ leads to Eq. (17).

\subsection{Estimation of parameters $\beta$ and $\gamma$}

Now, we define the model parameters $\beta$ and $\gamma$ by the requirement
that at the weak field limit our model is converted into the Heisenberg$-$Euler electrodynamics. Expanding Lagrangian (1) at small value $\beta{\cal F}\ll 1$, we have
\begin{equation}
{\cal L}=-{\cal F}+3\beta{\cal F}^2-\frac{15}{2}\beta^2{\cal F}^3+{\cal O}\left((\beta{\cal F})^4\right)+\frac{\gamma}{2}{\cal G}^2.
\label{24}
\end{equation}
The QED Lagrangian with one loop correction (the Heisenberg$-$Euler Lagrangian) is given by \cite{Gies}
\begin{equation}
{\cal L}_{HE}=-{\cal F}+c_1{\cal F}^2+c_2{\cal G}^2,~~~c_2=\frac{14\alpha^2}{45m_e^4},~~~c_1=\frac{8\alpha^2}{45m_e^4},
\label{25}
\end{equation}
where the coupling constant $\alpha=e^2/(4\pi)\approx 1/137$ and the electron mass $m_e=0.51~\mbox{MeV}$. By identifying Eqs. (24) and (25) we obtain
\begin{equation}
\beta=\frac{8\alpha^2}{135m_e^4}=4.6\times 10^{-5}~\mbox{MeV}^{-4},~~~\gamma=\frac{28\alpha^2}{45m_e^4}= 4.9\times 10^{-4}~\mbox{MeV}^{-4}.
\label{26}
\end{equation}
According to the Heisenberg$-$Euler observations it is possible scattering of photons
(light-by-light scattering) through a quantum-loop process involving virtual electron and positron pairs.
Equations (24)-(26) show that two-photon physics is the same in the framework of QED and in the NED (1).
But three-photon physics will be different. Experiments at ATLAS/LHC involve only two-photon physics,
and therefore, future three-photon experiments could distinguish QED from other NED models.

\section{GR coupled to NED}

The action of GR coupled to electromagnetic fields is given by
\begin{equation}
S=\int d^4x\sqrt{-g}\left[\frac{1}{2\kappa^2}R+ {\cal L}\right],
\label{27}
\end{equation}
where $\kappa^2=8\pi G$ with $G$ being the Newton constant. From action (27) one finds equations as follows:
\begin{equation}
R_{\mu\nu}-\frac{1}{2}g_{\mu\nu}R=-\kappa^2T_{\mu\nu},
\label{28}
\end{equation}
\begin{equation}
\partial_\mu\left(\frac{\sqrt{-g}F^{\mu\nu}(\beta{\cal F}-1)}{(2\beta{\cal F}+1)^{5/2}}\right)=0.
\label{29}
\end{equation}
The squared of the line element of homogeneous and isotropic spacetime is given by
\begin{equation}
ds^2=-dt^2+a(t)^2\left(dx^2+dy^2+dz^2\right),
\label{30}
\end{equation}
with $a(t)$ being a scale factor. We assume that the cosmic background filled by stochastic magnetic fields. The averaged magnetic fields, which guaranty the isotropy of the Friedman$-$Robertson$-$Walker (FRW) spacetime, obey the equations \cite{Tolman}
\begin{equation}
\langle\textbf{B}\rangle=0,~~~~\langle E_iB_j\rangle=0,~~~~\langle B_iB_j\rangle=\frac{1}{3}B^2g_{ij}.
\label{31}
\end{equation}
Here, the brackets $\langle.\rangle$ mean an average over a volume but for a simplicity in the following we will omit the brackets. The NED stress-energy tensor with Eq. (31) represents a perfect fluid \cite{Novello1}. Indeed, taking into account Eq. (31), we obtain from Eq. (2) at $\textbf{E}=0$ the components of the stress-energy tensor $T^{\mu\nu}=-\mbox{diag}(\rho,p,p,p)$ corresponding to a perfect fluid tensor, where the energy density and pressure are given by
\begin{equation}
\rho=-{\cal L}-E^2{\cal L}_{\cal F}+{\cal G}{\cal L}_{\cal G}=\frac{(1-\beta{\cal F})E^2}{(1+2\beta{\cal F})^{5/2}} +\frac{{\cal F}}{(1+2\beta{\cal F})^{3/2}}+\frac{1}{2}\gamma{\cal G}^2,
\label{32}
\end{equation}
\begin{equation}
p={\cal L}+\frac{E^2-2B^2}{3}{\cal L}_{\cal F}-{\cal G}{\cal L}_{\cal G}=-\frac{{\cal F}}{(2\beta{\cal F}+1)^{3/2}}+\frac{(E^2-2B^2)(\beta{\cal F}-1)}{3(2\beta{\cal F}+1)^{5/2}}-\frac{1}{2}\gamma{\cal G}^2.
\label{33}
\end{equation}
The Friedman equation is given as follows:
\begin{equation}
3\frac{\ddot{a}}{a}=-\frac{\kappa^2}{2}\left(\rho+3p\right),
\label{34}
\end{equation}
where $\dot{a}(t)=\partial a/\partial t$. When $\rho + 3p < 0$ we have $\ddot{a}>0$ and the universe acceleration holds. An isotropic symmetry is guarantied by the equation $\langle\textbf{B}\rangle = 0$. Making use of Eqs. (32) and (33) we find (in the case of $\textbf{E}=0$)
\begin{equation}
\rho+3p=-\frac{B^2(2\beta B^2-1)}{(1+\beta B^2)^{5/2}}.
\label{35}
\end{equation}
The plot of $\beta(\rho+3p)$ as a function of $\beta B^2$ is represented by Fig. 2.
\begin{figure}[h]
\includegraphics[height=4.0in,width=4.0in]{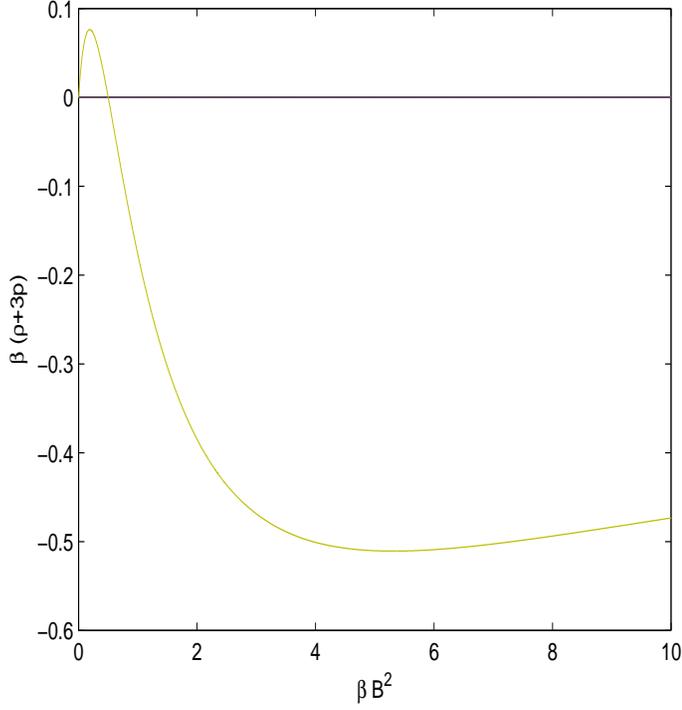}
\caption{\label{fig.2}The function  $\beta(\rho+3p)$ versus $\beta B^2$. }
\end{figure}
If $\beta B^2>0.5$ one has $\rho + 3p < 0$ and the universe acceleration occurs. As a result, the strong magnetic fields lead to the inflation of the universe. Consider the conservation of the stress-energy tensor, $\nabla^\mu T_{\mu\nu}=0$,
\begin{equation}
\dot{\rho}+3\frac{\dot{a}}{a}\left(\rho+p\right)=0.
\label{36}
\end{equation}
With the aid of Eqs. (32) and (33) if $\textbf{E} = 0$, one finds
\begin{equation}
\rho=\frac{B^2}{2\left(1+\beta B^2\right)^{3/2}},~~~~\rho+p=\frac{B^2(2-\beta B^2)}{3\left(1+\beta B^2\right)^{5/2}}.
\label{37}
\end{equation}
It follows from Eqs. (35) and (37) that the strong energy conditions $\rho+p\geq 0$, $\rho+3p\geq 0$, which include the null energy condition $\rho+p\geq 0$, take place only for $\beta B^2\leq 0.5$ corresponding to the deceleration phase. At the inflation phase, $\rho+3p<0$, the strong energy conditions are violated.
Taking into account Eq. (37), one finds the solution to Eq. (36)
\begin{equation}
B(t)=\frac{B_0}{a^2(t)},
\label{38}
\end{equation}
where $B_0$ is the value of the magnetic field which corresponds to $a(t)=1$.
The scale factor $a(t)$ increases due to the universe expansion and the magnetic field $B(t)$ decreases. With the help of Eqs. (37) and (38) one obtains
\begin{equation}
\rho(t)=\frac{a^2(t) B_0^2}{2\left(a^4(t)+\beta B_0^2\right)^{3/2}},
~~~~p(t)=\frac{a^2(t)B_0^2(a^4(t)-5\beta B_0^2)}{6\left(a^4(t)+\beta B_0^2\right)^{5/2}}.
\label{39}
\end{equation}
Making use of Eq. (39) we find
\begin{equation}
\lim_{a(t)\rightarrow 0}\rho(t)=\lim_{a(t)\rightarrow 0}p(t)=\lim_{a(t)\rightarrow \infty}\rho(t)=\lim_{a(t)\rightarrow \infty}p(t)=0.
\label{40}
\end{equation}
Equation (40) shows that there are not singularities of the density of the energy and pressure as $a(t)\rightarrow 0$ and $a(t)\rightarrow \infty$. The plot of the equation of state (EoS) $w=p(t)/\rho(t)$ versus $x=a(t)/(\beta B_0^2)^{1/4}$ is in Fig. 3.
\begin{figure}[h]
\includegraphics[height=4.0in,width=4.0in]{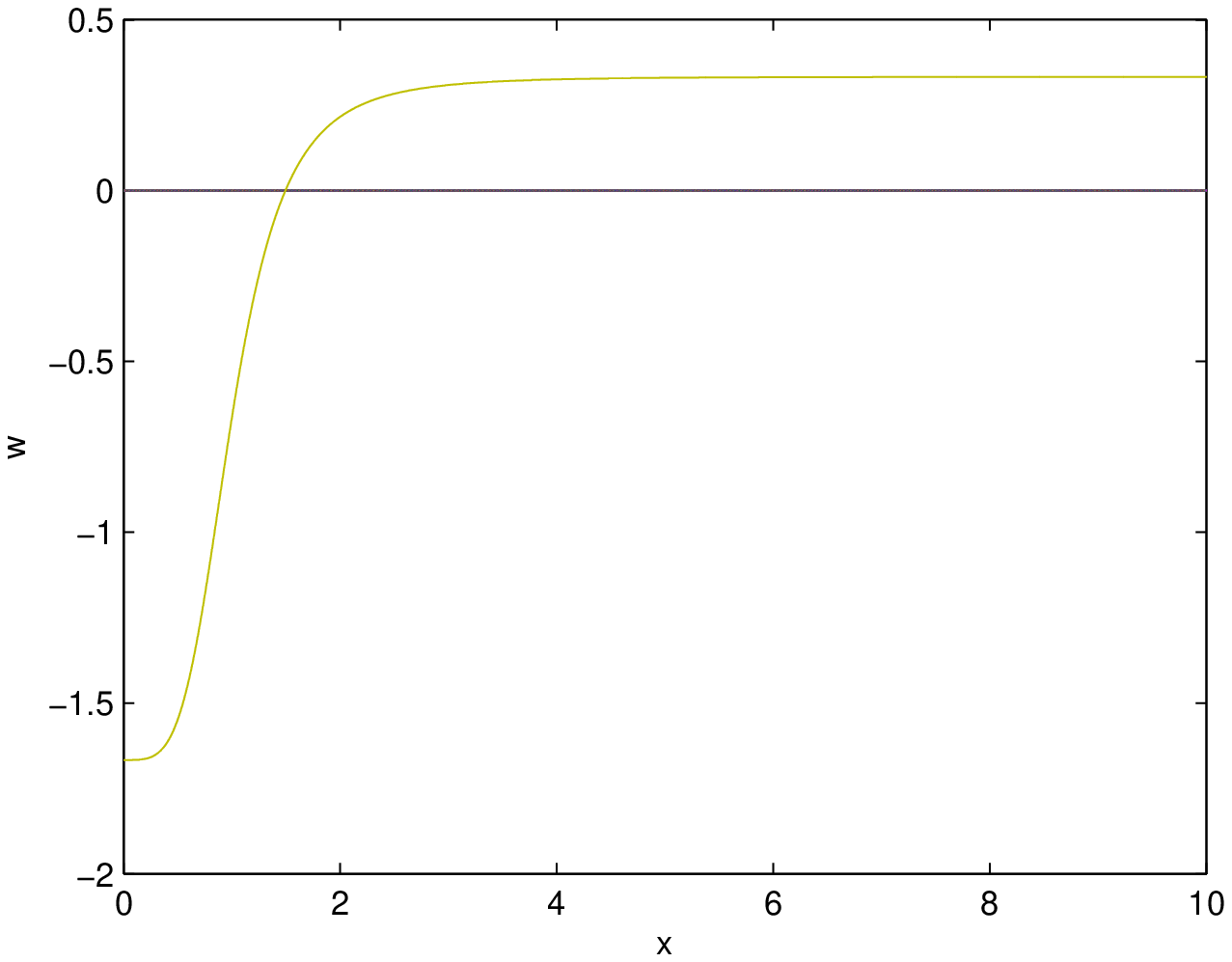}
\caption{\label{fig.3}The function  $w$ versus $x=a/(\beta B_0^2)^{1/4}$.}
\end{figure}
Making use of Eq. (39) one obtains
\begin{equation}
\lim_{x\rightarrow\infty} w=\lim_{x\rightarrow\infty}\frac{x^4-5}{3(x^4+1)}=\frac{1}{3}.
\label{41}
\end{equation}
In accordance with Eq. (41) the EoS corresponds to the ultra-relativistic behaviour \cite{Landau} as $a(t)\rightarrow \infty$.
The de Sitter spacetime, $w=-1$, is realized for $x=1/\sqrt[4]{2}\approx 0.84$. From Eq. (3), (28) and (38), we find the Ricci scalar
\begin{equation}
R=\kappa^2{\cal T}=\frac{3\kappa^2\beta B^4}{(1+\beta B^2)^{5/2}}=
\frac{3\kappa^2a^2(t)\beta B_0^4}{\left(a^4(t)+\beta B_0^2\right)^{5/2}}=
\kappa^2\left[\rho(t)-3p(t)\right].
\label{42}
\end{equation}
The $\beta R/\kappa^2$ as a function of $[1/(\beta B_0^2)]^{1/4}a$ is given by Fig. 4.
\begin{figure}[h]
\includegraphics[height=4.0in,width=4.0in]{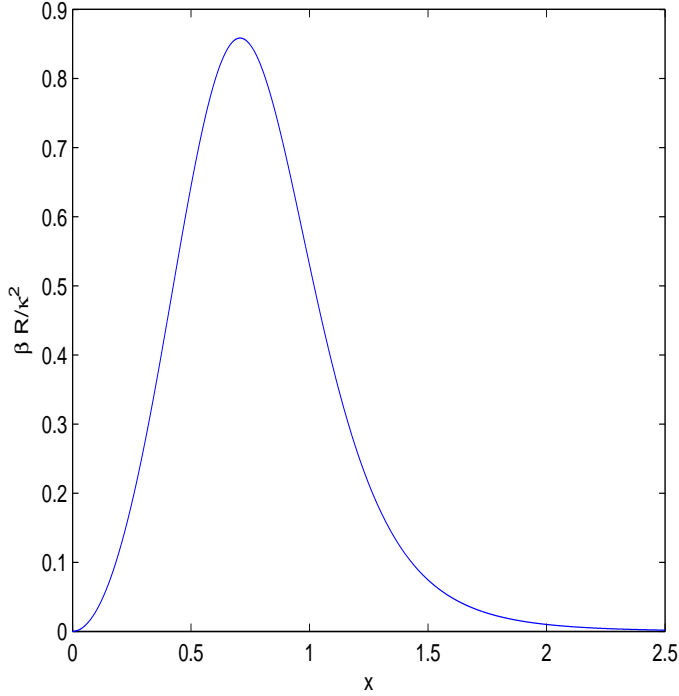}
\caption{\label{fig.4}The function  $\beta R/\kappa^2$ versus $x\equiv a/(\beta B_0^2)^{1/4}$. }
\end{figure}
With the help of Eqs. (40) and (42) one obtains
\begin{equation}
\lim_{a(t)\rightarrow 0}R(t)=\lim_{a(t)\rightarrow \infty}R(t)=0.
\label{43}
\end{equation}
In accordance with Eq. (43) there is not a singularity of the Ricci scalar. The Kretschmann scalar $R_{\mu\nu\alpha\beta}R^{\mu\nu\alpha\beta}$ and the Ricci tensor squared $R_{\mu\nu}R^{\mu\nu}$  can be expressed via $\kappa^4\rho^2$, $\kappa^4\rho p$, and $\kappa^4p^2$ \cite{Kruglov1}. Then they become zero as $a(t)\rightarrow 0$ and $a(t)\rightarrow \infty$.
During the evolution of the universe the scale factor increases as $t\rightarrow\infty$ and spacetime becomes flat. Equations (35) and (38) show that the universe acceleration takes place at $a(t)<(2\beta)^{1/4}\sqrt{B_0}\approx 1.19\beta^{1/4}\sqrt{B_0}$.

\section{The universe evolution}

For the three dimensional flat universe the second Friedman equation is
\begin{equation}
\left(\frac{\dot{a}}{a}\right)^2=\frac{\kappa^2\rho}{3}.
\label{44}
\end{equation}
By virtue of Eqs. (37), (38) and (44), one obtains
\begin{equation}
\dot{a} =\frac{\kappa B_0a^2}{\sqrt{6}(a^4+\beta B_0^2)^{3/4}}.
\label{45}
\end{equation}
Using the unitless variable $x=a/(\beta^{1/4}\sqrt{B_0})$, Eq. (45) is rewritten as
\begin{equation}
\dot{x} =\frac{\kappa x^2}{\sqrt{6\beta}(x^4+1)^{3/4}}.
\label{46}
\end{equation}
The function $y\equiv\sqrt{6\beta}\dot{x}/\kappa$ is given by Fig. 5.
\begin{figure}[h]
\includegraphics[height=4.0in,width=4.0in]{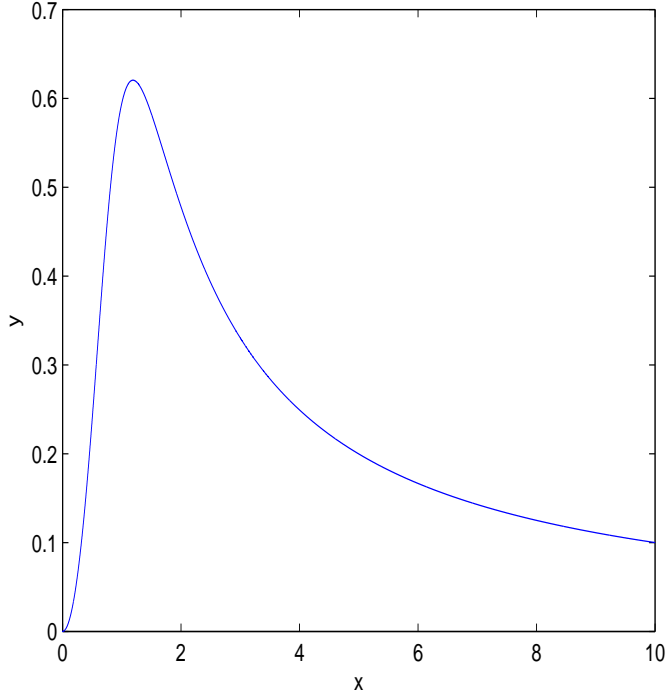}
\caption{\label{fig.5}The function $y\equiv\sqrt{6\beta}\dot{x}/\kappa$ versus $x$.}
\end{figure}
According to Fig. 5 at the initial time the universe inflation takes place, ($\dot{y}>0$), then the graceful exit occurs at the point $x=\sqrt[4]{2}$ ($\dot{y}=0$) and after the universe decelerates.
From Eq. (46) we obtain
\begin{equation}
\int_\epsilon^x \frac{(x^4+1)^{3/4}}{x^2}dx =\frac{\kappa}{\sqrt{6\beta}}\int_0^t dt.
\label{47}
\end{equation}
Calculating the integrals in Eq. (47) one arrives at
\[
x^3{_2}F_1\left(\frac{1}{4},\frac{3}{4};\frac{7}{4};-x^4\right)-\frac{(x^4+1)^{3/4}}{x}
\]
\begin{equation}
-\epsilon^3{_2}F_1\left(\frac{1}{4},\frac{3}{4};\frac{7}{4};-\epsilon^4\right)
+\frac{(\epsilon^4+1)^{3/4}}{\epsilon}=\frac{\kappa t}{\sqrt{6\beta}},
\label{48}
\end{equation}
${_2}F_1(a,b;c;z)$ is the hypergeometric function, $\epsilon$ be consistent with the starting of the universe inflation. We can study the evolution of the universe inflation from Eq. (48).
To obtain the asymptotic of the scale factor as $t\rightarrow\infty$ we explore the relation \cite{abram}
\[
{_2}F_1(a,b;c;z)=\frac{\Gamma(c)\Gamma(b-a)}{\Gamma(b)\Gamma(c-a)}(-z)^{-a}{_2}F_1(a,1-c+a;1-b+a;1/z)
\]
\begin{equation}
+\frac{\Gamma(c)\Gamma(a-b)}{\Gamma(a)\Gamma(c-b)}(-z)^{-b}{_2}F_1(b,1-c+b;1-a+b;1/z).
\label{49}
\end{equation}
Making use of Eq. (49) one obtains
\[
{_2}F_1(1/4,3/4;7/4;-x^4)=\frac{\Gamma(7/4)\Gamma (1/2)} {\Gamma(3/4)\Gamma(3/2)}(x)^{-1}{_2}F_1(1/4,-1/2;1/2;-1/x^4)
\]
\begin{equation}
+\frac{\Gamma(7/4)\Gamma(-1/2)}{\Gamma(1/4)\Gamma(1)}(x)^{-3}{_2}F_1(3/4,0;3/2;-1/x^4).
\label{50}
\end{equation}
Expanding the hypergeometric functions in $1/x^4\rightarrow 0$ in the leading order, we find from Eq. (48) $1.66x^2\approx \kappa t/\sqrt{6\beta}$ and the scale factor is given by
\begin{equation}
a(t)\approx 0.5\sqrt{\kappa B_0 t}~~~~~t\rightarrow \infty.
\label{51}
\end{equation}
Equation (51) shows that the behavior of the scale factor as $t\rightarrow\infty$ corresponds to the radiation era. The similar behavior of the scale factor occurs in another model proposed in \cite{Kruglov4} where as $t\rightarrow \infty$ we have $a(t)\rightarrow (2/3)^{1/4}\sqrt{\kappa B_0 t}\approx 0.9\sqrt{\kappa B_0 t}$. At the same time at the earlier time corresponding to the inflation the evolution of the scale factor is different.

Let us consider the deceleration parameter, making use of Eqs. (34), (37), (38) and (44),
\begin{equation}
q=-\frac{\ddot{a}a}{(\dot{a})^2}=\frac{x^4-2}{x^4+1}.
\label{52}
\end{equation}
\begin{figure}[h]
\includegraphics[height=4.0in,width=4.0in]{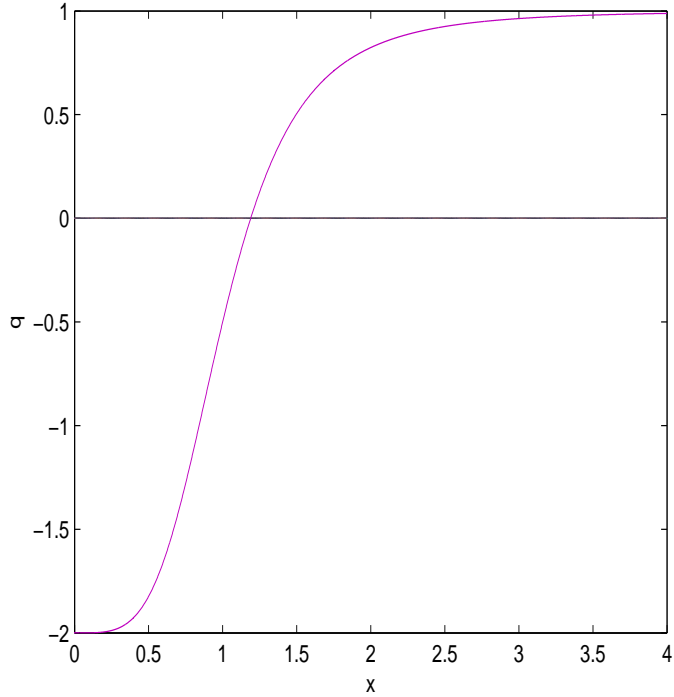}
\caption{\label{fig.6}The function $q$ versus $x=a/(\beta B_0^2)^{1/4}$.}
\end{figure}
In Fig. (6) the deceleration parameter $q$ versus $x=a/(\beta B_0^2)^{1/4}$ is depicted. The inflation ($q<0$) occurs until the graceful exit point $x=\sqrt[4]{2}\approx 1.19$.
The deceleration parameter becomes zero at $x=\sqrt[4]{2}$ and then the deceleration phase ($q>0$) takes place. 
The deceleration parameter in the model \cite{Kruglov4} is given by $q=(2x^4-3)/(2x^4+1)$ ($x=a/(\beta B_0^2)^{1/4}$) and begins at the very small value $q=-3$ ($a(t)=0$) compared with the value $q=-2$ in the present model. In the model \cite{Kruglov5} $q=-5$ ($a(t)=0$). But the expansion of the universe has to be closer to the de Sitter type with $q=-1$. Presently there is no evidence for such super-inflationary behavior as in \cite{Kruglov4}, \cite{Kruglov5}.

Now we estimate the amount of the inflation using the e-folding \cite{Liddle}
\begin{equation}
N=\ln\frac{a(t_{end})}{a(t_{in})},
\label{53}
\end{equation}
where $t_{in}$ is an initial time and $t_{end}$ is the final time of the inflation. Using the graceful exit point $x\approx 1.19$ one obtains $a(t_{end})\approx 1.19 b$ ($b\equiv \beta^{1/4}\sqrt{B_0}$).
It is known that the horizon and flatness problems may be solved when e-folding is $N\approx 70$ \cite{Liddle}. From Eq. (53) we obtain the scale factor corresponding to the initial time of the inflation
\begin{equation}
a(t_{in})=\frac{1.19b}{\exp(70)}\approx 4.7\times 10^{-31}b.
\label{54}
\end{equation}
Then $\epsilon\approx 4.7\times 10^{-31}$.
We use the units $\kappa=\sqrt{8\pi G}=4.1\times 10^{-28}~\mbox{eV}^{-1}$, $\beta=4.6\times 10^{-29}~\mbox{eV}^{-4}$, $1~\mbox{s}=1.5\times 10^{15}~\mbox{eV}^{-1}$
to calculate the duration of the inflationary period. Then one obtains $\kappa/\sqrt{6\beta}=2.47\times10^{-14}\mbox{eV}=37~\mbox{s}^{-1}$.
Using the value $x=\sqrt[4]{2}\approx 1.19$, which corresponds to the end of the inflation, the duration of the universe inflation, according to Eq. (48), will be large. As a result, the universe inflation will be eternal. Therefore, it is difficult to describe other epochs. If we use the time duration $1~\mbox{s}$, Eq. (47) results the value $\epsilon=0.0267$ for $x\approx 1.19$. Then the e-folding number (53) becomes $N\approx 3.8$ that is small for the resolution of the horizon and flatness problems. One can vary the initial time $\epsilon$ and to see the behaviour of the universe inflation, the e-folding number, and the duration of the inflation. Thus, there are phases of the universe acceleration, deceleration and the graceful exit that are the attractive property of the model under consideration. The models \cite{Kruglov4}, \cite{Kruglov5} also possess the similar phases.

\section{Classical stability, causality and the speed of sound}

The causality takes place when the speed of the sound is less than the local light speed, $c_s\leq 1$ \cite{Quiros}. If the square sound speed is positive ($c^2_s> 0$) a classical stability holds. From Eqs. (32) and (33) one can find the sound speed squared (for the case of $E=0$)
\begin{equation}
c^2_s=\frac{dp}{d\rho}=\frac{5\beta^2 B^4-23\beta B^2+2)}{3(\beta B^2+1)(2-\beta B^2)}.
\label{55}
\end{equation}
The classical stability occurs when $c^2_s> 0$, and gives the bound
\begin{equation}
0<\beta B^2<\frac{23-\sqrt{489}}{10}\approx 0.09~~\mbox{or}~~2<\beta B^2<\frac{23+\sqrt{489}}{10}\approx 4.5.
\label{56}
\end{equation}
The causality $c^2_s\leq 1$ requires that
\begin{equation}
0\leq\beta B^2<2~~\mbox{or}~~\beta B^2\geq \frac{13+\sqrt{201}}{8}\approx 3.4.
\label{57}
\end{equation}
Equations (56) and (57) give the bounds
\begin{equation}
0\leq\beta B^2<\frac{23-\sqrt{489}}{10}~~~ \mbox{or}~ \frac{13+\sqrt{201}}{8}\leq\beta B^2<\frac{23+\sqrt{489}}{10}.
\label{58}
\end{equation}
The principles of causality and unitarity, studied in Sec. 2, require $\beta B^2\leq(11-\sqrt{105})/4\approx0.19$. This restriction and those in Eqs. (58) take place when the inequality is satisfied at the deceleration phase of the universe evolution:
\begin{equation}
0\leq\beta B^2<\frac{23-\sqrt{489}}{10}\approx 0.09.
\label{59}
\end{equation}
The acceleration phase is realized at $\beta B^2>0.5$ and the classical stability, causality and unitarity are violated in this phase. The singularity that appears in the speed of sound at $\beta B^2=2$ belongs to the acceleration phase where the classical stability, causality and unitarity are broken.

\section{The spectral index, tensor-to-scalar ratio, and the running of the spectral index}

From Eqs. (32) and (33) at $\textbf{E}=0$ one finds the equations
\begin{equation}
p=-\rho+\frac{2\rho(2-\beta B^2)}{3(\beta B^2+1)},
\label{60}
\end{equation}
\begin{equation}\label{61}
 4\rho^2(\beta B^2+1)^3- B^4=0.
\end{equation}
One can obtain the solution (for $\beta B^2$ as a function of $\beta\rho$) to the cubic equation (61) and to place it to Eq. (60) obtaining the equation of state for perfect fluid
\begin{equation}
p=-\rho+f(\rho).
\label{62}
\end{equation}
When the condition $|f(\rho)/\rho|\ll 1$ is satisfied the expressions for the spectral index $n_s$, the tensor-to-scalar ratio $r$, and the running of the spectral index $\alpha_s=dn_s/d\ln k$ are  \cite{Odintsov}
\begin{equation}
n_s\approx 1-6\frac{f(\rho)}{\rho},~~~r\approx 24\frac{f(\rho)}{\rho},~~~\alpha_s\approx -9\left(\frac{f(\rho)}{\rho}\right)^2.
\label{63}
\end{equation}
From Eq. (63) we find the equation
\begin{equation}
r=4(1-n_s)=8\sqrt{-\alpha_s}.
\label{64}
\end{equation}
It worth mentioning that the relation (64) holds also for the models \cite{Kruglov4} and \cite{Kruglov5}.
In accordance with the PLANCK experiment \cite{Ade} and WMAP data \cite{Komatsu}, \cite{Hinshaw} we have
\[
n_s=0.9603\pm 0.0073 ~(68\% CL),~~~r<0.11 ~(95\%CL),
\]
\begin{equation}
\alpha_s=-0.0134\pm0.0090 ~(68\% CL).
\label{65}
\end{equation}
When we take $r=0.13$, using Eq. (65), the values for the spectral index is $n_s=0.9675$ and the running of the spectral index is $\alpha_s=-2.64\times 10^{-4}$. Using Eq. (63) one obtains the value $f(\rho)/\rho\approx 0.005$ and from Eq. (60) the relation is
\begin{equation}
\frac{f(\rho)}{\rho}=\frac{2(2-\beta B^2)}{3(\beta B^2+1)}.
\label{66}
\end{equation}
In the models \cite{Kruglov4} and \cite{Kruglov5} the expressions $f(\rho)/\rho$ are different.
Then from Eq. (66) we find the value (for $r=0.13$) of the magnetic field $B\approx 1.4/\sqrt{\beta}\approx 206~\mbox{MeV}^2\approx 10^{12}$ T that corresponds to the inflation phase with the maximum of the energy density $\beta\rho\approx 0.192$.

\section{Conclusion}

We have considered a NED model where a singularity of the electric field in the center of charges is absent similar to BI electrodynamics. The principles of causality, the classical stability and unitarity were investigated. The interval of electromagnetic fields when causality, the classical stability and unitarity take place, were obtained. The dual symmetry is violated in this model because of dimensional parameters $\beta$ and $\gamma$. It was shown that corrections to Coulomb's law are in the order of ${\cal O}(r^{-6})$. The magnetic universe with a stochastic background $\langle B^2\rangle \neq 0$ was studied and we demonstrated that the model with homogeneous and isotropic cosmology explains the universe inflation. There are no singularities of the energy density, pressure, the Ricci scalar, the Ricci tensor squared, and the Kretschmann scalar in our model. A stochastic magnetic field is the source of the inflation of the universe. At $B< 1/\sqrt{2\beta}$ the universe decelerates leading to the radiation era. The spectral index, the tensor-to-scalar ratio, and the running of the spectral index calculated are roughly in accord with the PLANCK and WMAP data. The attractive feature in our model of inflation that there is the graceful exit. There are similarities and differences of the current model with \cite{Kruglov4} and \cite{Kruglov5}. The general behaviour of the evolution of the universe inflation in these models are similar. The values of the background magnetic field when the principles of causality, the classical stability and unitarity hold, the corrections to Coulomb law, the ending point of inflation, when the the PLANCK experiment and WMAP data in agreement with the model prediction, are different. Also, this model avoids the super-inflationary behavior as in \cite{Kruglov4} and \cite{Kruglov5}.

We summarise the results obtained as follows. The particular structure of the NLED Lagrangian (1) leads to attractive model features: the causality and unitarity of the model take place for some range of electromagnetic fields, the existence of the acceleration, deceleration phases and graceful exit of the universe inflation, the absence of curvature singularities, the EoS corresponds to the ultra-relativistic behaviour as $a(t)$ goes to infinity (and the absence  super-inflationary behavior), cosmological parameters are approximately in agreement with the PLANCK and WMAP data.

\end{document}